\documentstyle[preprint,prb,eqsecnum,aps]{revtex}

\begin{document}
\draft
\preprint{}
\title{Phase String Effect in the $t-J$ Model: General Theory}
\author{Z. Y. Weng, D. N. Sheng, Y. -C. Chen,$^*$ and C. S. Ting }
\address{Texas Center for Superconductivity and Department of Physics\\
University of Houston, Houston, TX 77204-5506 }
\maketitle
\begin{abstract}
We reexamine the problem of a hole moving in an antiferromagnetic spin 
background and find that the injected hole will always pick up a sequence of nontrivial
phases from the spin degrees of freedom. Previously unnoticed, such a string-like phase 
originates from the hidden Marshall signs which are scrambled by the hopping of
the hole.  We can rigorously show that this phase string is  non-repairable at low energy
and give a general proof  that the  spectral weight $Z$ must vanish at the ground-state energy
due to the phase string effect. Thus, the quasiparticle description fails here and the quantum 
interference effect  of the phase string dramatically affects the long-distance behavior of the
injected hole. We introduce a so-called phase-string  formulation of
the $t-J$ model for a general number of holes in which the phase string effect 
can be  explicitly tracked.  As an example, by applying this new mathematical formulation in one dimension, we reproduce the well-known Luttinger-liquid behaviors of the  asymptotic single-electron  Green's function and the spin-spin correlation function. We can also use the present phase string 
theory  to justify previously developed spin-charge separation theory  in two dimensions, which 
offers a systematic explanation for the transport and magnetic anomalies in the high-$T_c$ cuprates.

\end{abstract}
\vspace{0.4in}
\pacs{71.27.+a, 74.20.Mn, 74.72.-h, 75.10.Jm}  

\narrowtext

\newpage

\section{INTRODUCTION}

A general interest in the $t-J$ model is motivated by the following experimental facts 
in the high-$T_c$ cuprates: an antiferromagnetic (AF) long-range order of $Cu$ spins 
in the $CuO_2$ layers exists in the insulating phase and a metallic phase emerges after 
the doped holes destroys the magnetic ordering,  where superconducting condensation
as well as anomalous normal-state properties are found. The $t-J$ model is composed 
of two terms: $H_{t-J}=H_J+H_t$, where  $H_J$  describes the AF superexchange 
coupling between the nearest-neighboring spins (as defined in (\ref{e01})) which 
fully explains the magnetic insulating phase in the cuprates, and $H_t$  describes 
the hopping of holes on such a spin background (as defined in (\ref{ehop})). The 
highly nontrivial  competition between the superexchange and hopping processes in the 
$t-J$ model  generates strong correlations among electrons, and is believed 
by many people to be the key to explain the strange-metal behaviors in the cuprates. 
Even though such a model has been  intensively studied for many years,  very few 
properties have been {\it reliably} understood  in the two dimensional (2D) doped  case, 
which  is presumably relevant to the metallic phase  of the high-$T_c$ cuprates.

To see the difficulty involved in this problem, let us take as an example one of the simplest 
cases: only one hole is present in the AF spin background. The motion of the hole 
usually creates a spin mismatch along its path.\cite{dagotto} Namely, the hopping changes 
the spin configuration, which otherwise would have perfect antiferromagnetic correlations.
For the N\'{e}el order, such a ``string''-like spin mismatch is easy to see,\cite{trugman} but it is not uniquely restricted to the case with a long-range order.   
It has been realized that such a spin mismatch left on the spin background
by the mobile hole could cost an energy
linearly proportional to its length and thus has to be ``repaired'' in order 
to allow the hole to move around freely. In fact, it has been well-known\cite{lee} that
a spin flip process can ``repair''  the spin mismatch. With the 
spin mismatch generated by the hopping being repairable by spin flips, the doped hole is 
generally believed to be a mobile object.

However, the crucial issue\cite{anderson} is whether such a mobile
hole can be described as quasiparticle characterized by a nonzero
spectral weight $Z$. Physically, a 
finite $Z$ implies that the hole  
only carries a local spin distortion (``spin-polaron'') as it moves. This is
a picture familiar in a conventional metal, where a spin-polaron is
usually replaced by, say, a phonon-polaron. Here a  spin-polaron
picture can indeed be  obtained by a self-consistent Born approximation,\cite{lee,born} which
is also supported by the finite-size exact diagonalization calculations.\cite{dagotto,num} 
However, different from the
phonon-polaron picture, $SU(2)$ spins are involved here and a $U(1)$ phase
may play an important role in shaping the long-distance part of the  
spin-polaron with {\it little} energy cost.\cite{remark1,shraiman} The question whether the spectral weight 
$Z$ vanishes at the  ground-state energy is particularly sensitive to such
long-wavelength, low-energy effects. Self-consistent perturbative approaches and 
numerical calculations themselves cannot provide a definite answer for it. 
In fact, Anderson\cite{anderson} has given a general argument that $Z$
has to vanish due to the existence of the upper-Hubbard band. A vanishing $Z$
means that each hole added to the system will    
cause a global change in the ground state,  and thus the resulting state
cannot be simply described as a quasi-particle-type excitation and
treated perturbatively. 

Thus a more accurate description of the long-distance effect is needed
in the present system in order to resolve this issue. As  the spin-mismatch 
left on the 
spin background  has to be restored  to avoid a linear  
potential energy, one would expect the quasiparticle picture to 
be generally correct, {\it 
unless the hole  picks up a nontrivial phase at each hopping step}. The
quantum interference effect of such a phase sequence, if  the latter exists,
can then dramatically change the long-wavelength behavior of the hole, leading to 
non-quasiparticle-like 
properties. In fact, in the one-dimensional (1D) case such a $U(1)$ phase 
string has already been demonstrated,\cite{weng1} where it plays a crucial role in 
shaping a non-Fermi-liquid (i.e., Luttinger liquid) behavior.  In the 
present paper, we will rigorously demonstrate  that for general 
dimensionality the injected
hole always has to pick up a sequence of $U(1)$ phases from the spin
background when it moves around, and the resulting phase is not repairable at low 
energy in contrast to the afore-mentioned repairable spin-mismatch string.  This phase string  
depends on  the instant spin configuration encountered by the hole and can be determined 
by a simple counting. We will then be able to prove\cite{sheng} that such a phase
string effect leads to a vanishing spectral weight $Z$ at the ground state 
energy in low dimensions. 

This nontrivial effect of the $U(1)$ phase string at large distances is 
generally present even when there are many 
doped holes,  regardless of whether the ground state possesses an AF 
long-range order or not. Therefore, the phase string effect is expected to be 
the most crucial factor in determining the low-energy, long-wavelength physics 
both for the one-hole problem and the finite doping case. A perturbative method, 
which may well describe the spin-polaron effect surrounding the doped hole, 
usually fails to account for this string effect. This is because the phase string effect is 
basically a nonlocal effect, but conventional approximations usually average out the 
effect locally  and thus result in a serious  problem at a long distance. The natural way
to avoid this difficulty is to find
a method for accurately tracking the phase string effect at large scales.  We will 
show that such a nonlocal effect can be explicitly ``counted'' by introducing 
``mutual statistics'' between  spins and holes. In fact, one can exactly 
map the phase string effect to a 
statistics-transmutation problem. The latter can be further transformed into
a nonlocal interacting problem if one recalls that statistics-transmutation
can be realized by a composite-particle representation,\cite{wilczek,laughlin} with the 
underlying particle with conventional statistics bound to a flux-tube. This is an exact 
reformulation of the $t-J$ model, which is of course mathematically equivalent 
to the conventional slave-particle representations. It has an advantage over 
the other formalisms, however,   due to the fact that  the nonlocal 
phase string effect hidden in the 
original Hamiltonian is now made explicitly, so that its long-distance effects 
can be tracked even after making a local  approximation in the Hamiltonian. 

One dimensional case can serve as a direct test of the phase string effect. As
an example, the asymptotic single-electron Green's function and the spin-spin 
correlation function are calculated based on the phase-string formulation 
developed in this paper, and the well-known Luttinger-liquid behavior in this system is 
reproduced. This shows that the phase string is indeed essential in shaping the 
long-wavelength, long-time correlations. In the 2D case, as an example of the
phase string effect, a spin-charge separation theory previously developed based 
on the slave-boson formalism\cite{weng2} will be reproduced in the present formalism. A key feature involved in 
this theory is nonlocal interactions between the spin and charge degrees of
freedom as mediated by the Chern-Simons type gauge fields. We show that they
arise as a consequence of  the nonlocal phase string effect in 2D, and  the present phase-string formalism provides both physical and mathematical justification for these topological gauge fields, which have been shown\cite{weng2} to be responsible for  
anomalous transport and magnetic properties closely resembling the experimental 
features found in the high-$T_c$ cuprates. 

The remainder of the paper is organized as follows. In the next section, we 
discuss the phase string effect in the one-hole case, and show that it can lead
to vanishing spectral weight at the ground-state energy. In Sec. III, we
consider a many-hole case and introduce a new mathematical 
representation to  explicitly track  the phase string effect. As an 
example of application of this phase-string formulation, in Sec. III,  
we first calculate the asymptotic single-electron Green's function and the spin-spin 
correlation function for the 1D case. Then we re-derive a mean-field type picture for the 2D case
in which the phase string effect plays a central role in 
shaping transport and magnetic properties.  Finally, a summary is presented in Sec. V. 
 
\section{PHASE STRING EFFECT: ONE-HOLE CASE}

\subsection{Marshall Sign Rule}

Let us start with the undoped case. It is described by the  
superexchange Hamiltonian 
\begin{equation}\label{e01} 
H_J= J\sum_{\langle ij\rangle} \left[{\bf S}_i\cdot {\bf S}_j -\frac {n_in_j}{4}\right], 
\end{equation} 
which is equivalent to the Heisenberg model as the electron 
occupation  numbers $n_i=n_j=1$. According to Marshall\cite{marshall}, the 
ground-state wavefunction of the Heisenberg Hamiltonian for a bipartite lattice 
is real and satisfies a sign rule. This sign rule
requires that flips of two antiparallel spins at nearest-neighbor 
sites are always accompanied by a sign change in the wave 
function: i.e., $\uparrow\downarrow \rightarrow (-1)\downarrow\uparrow$. 
The Marshall sign rule may be easily understood as below. Suppose that 
one has a complete set of spin bases 
$\{|\phi\rangle\}$ with the built-in Marshall sign. It is straightforward
to verify that matrix elements of $H_J$ become negative-definite: 
$\langle\phi'|H_J|\phi\rangle\leq 0$. Then, for the ground state $|\psi_0\rangle=\sum_{\phi}\chi_{\phi}
|\phi\rangle$ one finds that the coefficient $\chi_{\phi}$ should always be real and 
positive (except for a trivial global phase) in order to reach the lowest 
energy.  It means that the Marshall sign is indeed the only sign present in 
the ground state. Marshall sign rule may even be approximately correct in some
other spin models with various frustrations.\cite{marshall2}

There are many ways to incorporate Marshall sign into the 
$S^z$-spin representation. 
We may divide a bipartite lattice into odd ($A$)  
and even ($B$) sublattices and assign an extra sign $-1$ to every down spin at
$A$ site. In this way, flips of two nearest-neighboring antiparallel spins always involve a down spin changing sublattices, and thus a sign change. 
This spin basis may be written as 
\begin{equation}\label{} 
|\phi\rangle= (-1)^{N_A^{\downarrow}} |\uparrow ...\downarrow\uparrow ...\downarrow\rangle  , 
\end{equation} 
where $N_A^{\downarrow}$ denotes the total number of down spins 
at A-sublattice. A matrix element of $H_J$  under the basis $\{|\phi\rangle\}$ 
satisfies
\begin{equation}
\langle\phi'|H_J|\phi\rangle \leq 0,
\end{equation}
for any spin configurations $\{\phi\}$ and $\{\phi'\}$. The above definition 
can be even generalized to the doped case. With the presence of
one hole, one may simply define the spin basis as 
\begin{equation}\label{esh} 
|\phi;(n)\rangle= (-1)^{N_A^{\downarrow}} |\uparrow ...\downarrow\uparrow\circ ...\downarrow\rangle ,
\end{equation} 
with $n$ denoting the hole site. 
It is easy to check that 
\begin{equation}\label{e2} 
\langle\phi';(n)|H_J|\phi;(n)\rangle \leq 0.
\end{equation}
This means that the Marshall sign rule is still satisfied when the hole is
fixed at a given site $n$.  

Now we consider the hopping of the hole. The hopping process is governed by  
$H_t$ term in the $t-J$ model which is defined by 
\begin{equation}\label{ehop} 
H_t= -t \sum_{\langle ij\rangle} c_{i\sigma}^{\dagger}c_{j\sigma} + H.c.  , 
\end{equation} 
where the Hilbert space is restricted by the no-double-occupancy constraint  
$\sum_{\sigma}c_{i\sigma}^{\dagger}c_{i\sigma}\leq 1$. Suppose that the hole 
initially at site $n$ hops onto a nearest-neighbor site $m$. The  
corresponding matrix element in the basis (\ref{esh}) is easily found
to be:  
\begin{equation}\label{e3}  
\langle\phi;(m)|H_t|\phi;(n)\rangle= - t  \sigma_m ,  
\end{equation}  
where $\sigma_m$ is the site-$m$ spin index in the state $|\phi;(n)\rangle$, and $|\phi;(m)\rangle$ 
is different from $|\phi; (n)\rangle$ by an exchange of the spin $\sigma_m$ with the hole
at site $n$. Since $\sigma_m=\pm 1$, the hopping matrix element is not 
sign-definite. In other words, the hopping process will lead to the violation 
of the Marshall sign rule in the ground state. In the following, we 
explore in detail this phase ``frustration'' effect introduced by the hopping of an injected hole. 

\subsection{Single-hole Green's function}

Starting with the ground state $|\psi_0\rangle=\sum_{\phi}\chi_{\phi}|\phi\rangle$ at 
half-filling, one can create a ``bare''  hole by removing away an electron in 
terms of the electron
operator $c_{i\sigma}$
\begin{equation}
c_{i\sigma}|\psi_0\rangle=(\sigma)^i\sum_{\phi}\chi_{\phi}|\phi; (i)\rangle. 
\end{equation} 
Here $\chi_{\phi}\geq 0$ and the sign $(\sigma)^i$ is from the Marshall sign originally assigned
to the spin $\sigma$ at the site $i$ as follows: 
if $\sigma=+1$, $(\sigma)^i=1$ and if $\sigma=-1$, $(\sigma)^i=(-1)^i=-1$ at
 A-sublattice site and $+1$ at B-sublattice site. 

One can track the evolution of such a bare hole by studying the propagator 
\begin{equation}\label{eprop1}
G_{\sigma}(j,i;  
E)=\langle\psi_0|c_{j\sigma}^{\dagger}G(E)c_{i\sigma}|\psi_0\rangle,
\end{equation}
with
\begin{equation}
G(E)=\frac 1 {E-H_{t-J}+i0^+}.
\end{equation}
By using the following  expansion in terms of $H_t$
\begin{equation}
G(E)= G_J(E)+ G_J(E)H_tG_J(E)+G_J(E)H_tG_J(E)H_tG_J(E)+...,
\end{equation}
with
\begin{equation}
G_J(E)=\frac 1{E-H_J+i0^+},
\end{equation}
$G_{\sigma}(j,i;  E)$ can be rewritten as
\begin{equation}
G_{\sigma}(j,i;  E)=(\sigma)^{j-i}\sum_{\phi',\phi}\chi_{\phi'}\chi_{\phi}\sum_{n=0}^{\infty}
\langle\phi'; (j)|G_J(E)\left(H_tG_J(E)\right)^n|\phi; (i)\rangle.
\end{equation}
Then we insert the following complete set of the basis states (\ref{esh}) into
the above expansion:
\begin{equation}
\sum_{m}\sum_{\{\phi\}}|\phi; (m)\rangle\langle\phi;(m)|=1.
\end{equation} 
By using the matrix element
(\ref{e3}) for the nearest-neighboring hopping, we further express the 
single-hole Green's function as follows
\begin{eqnarray}  
G_{\sigma}(j,i; & E)&= (\sigma)^{j-i}\sum_{\mbox{(all paths)}}\sum_{\mbox{(all  states)}} \chi_{\phi'} 
\chi_{\phi} \nonumber\\ & & \times T_{ij}^{path}\cdot 
\prod_{s=0}^{K_{ij}}\langle\phi^{s+1};{(m_s)}|G_J(E)|\phi^s;{(m_s)}\rangle, 
\end{eqnarray} 
where intermediate states $|\phi^s;{(m_s)}\rangle$ and $|\phi^{s+1};{(m_s)}\rangle$ 
describe two different spin configurations $\{\phi^s\}$ and $\{\phi^{s+1}\}$
with the hole sitting at site $m_s$ on a given path connecting sites $i$ and 
$j$: $m_0=i, m_1, ..., m_{K_{ij}}=j, $ (Here $K_{ij}$ is the total number of  
links for the given path, and $\phi^0\equiv \phi$,
$\phi^{K_{ij}+1}\equiv \phi'$). $T_{ij}^{path}$ is a product of 
matrices of $H_t$ which connects $\{|\phi^{s+1};{(m_s)}\rangle\}$ with $\{|
\phi^{s+1};{(m_{s+1})}\rangle\}$ for such a path: 
\begin{equation}\label{estring} 
T_{ij}^{path}=\prod_{s=1}^{K_{ij}}(-t)\sigma_{
m_s},
\end{equation}
where $\sigma_{m_s}$ denotes the instant spin state at site $m_s$ right before
the hole hops to it. 

We can further write $G_{\sigma}(j,i; E)$  
in a more compact form, namely,  
\begin{equation}\label{epath} 
G_{\sigma}(j,i; E)=-(\sigma)^{j-i}\sum_{\mbox{(all paths)}}\sum_{\{\tilde{\phi}\}}W_{path}[\{\tilde{\phi}\}] 
\left(\prod_{s=1}^{K_{ij}}\sigma_{m_s}\right), 
\end{equation} 
where the summation over $\{\tilde{\phi}\}$ means summing over all the possible spin
configurations in the initial and final, as well as the intermediate states. Here 
$W_{path}[\{\tilde{\phi}\}]$ is defined by
\begin{equation}\label{ew}
W_{path}[\{\tilde{\phi}\}]\equiv \frac 1 t \chi_{\phi'} \chi_{\phi}  
\prod_{s=0}^{K_{ij}}(-t)\langle\phi^{s+1};{(m_s)}|G_J(E)|\phi^s;{(m_s)}\rangle.
\end{equation}

In the following, we prove that $W_{path}[\{\tilde{\phi}\}]$ is always positive-definite near the ground-state energy.
To determine the sign of 
$\langle\phi^{s+1};{(m_s)}|G_J(E)|\phi^s;{(m_s)}\rangle$, one may expand $G_J$ as follows 
\begin{equation} 
G_J(E)=\frac 1 E \sum_{n=0}^{\infty} \frac {{H_J}^n}{E^n}. 
\end{equation} 
Note that $\langle\phi^{s+1};(m_s)|{H_J}^n|\phi^s;{(m_s)}\rangle= (-1)^n 
|\langle\phi^{s+1};(m_s)|{H_J}^n|\phi^s;{(m_s)}\rangle| $ (one may easily show it by 
writing ${H_J}^n=H_J\cdot H_J\cdot ...$ and inserting the complete set of 
(\ref{esh}) in between and using the condition (\ref{e2})).  Then
one finds  
\begin{equation}\label{eg0} 
\langle\phi^{s+1};{(m_s)}|G_J(E)|\phi^s;{(m_s)}\rangle=\frac 1 E \sum_k \frac { 
|\langle\phi^{s+1}{(m_s)}|{H_J}^n|\phi^{s+1};{(m_s)}\rangle|}{(-E)^n} <0 , 
\end{equation} 
if $E<0$. Of course one still needs to determine the convergence range of the 
expansion. By inserting a complete set of eigenstates of $H_J$ (denoted as 
$\{|M; (m_s)\rangle\}$) as
intermediate states,  $\langle\phi^{s+1};{(m_s)}|G_J(E)|\phi^s;{(m_s)}\rangle$
can be also written in the form
\begin{equation}
\langle\phi^{s+1};{(m_s)}|G_J(E)|\phi^s;{(m_s)}\rangle=\sum_M\frac{
\langle\phi^{s+1};{(m_s)}|M; (m_s)\rangle \langle M; 
(m_s)|\phi^s;{(m_s)}\rangle}{E-E^0_M+i0^+}
\end{equation}
which is an analytic function of $E$ except for a branch cut on the real axis 
covered by the eigenvalues $\{E^0_M\}$ of $H_J$ (with a hole fixed at site $m_s$). This analytic 
property  will  guarantee the convergence of the expansion (\ref{eg0}) in the 
whole region of $E<E_G^0<0$ on the real axis, where $E_G^0$ is the 
lowest-energy eigenvalue of $H_J$ with the hole fixed on a lattice site. We note that  $E_G^0$ is always higher than the true ground-state energy 
$E_G$ of $H_{t-J}$, where the hole is allowed to move around to gain its kinetic energy. Therefore, near the ground-state energy 
$E_G$, one always has $W_{path}[\{\tilde{\phi}\}]\geq 0$.

Since $W_{path}[\{\tilde{\phi}\}]$ is sign-definite, one may introduce the following weight-functional for each path and an arbitrary 
spin configuration $\{\tilde{\phi}\} $: 
\begin{equation}\label{eweight}
\rho_{path}[\{\tilde{\phi}\}]=\frac {W_{path}[\{\tilde{\phi}\}]}{\sum_{\mbox{(all paths)}}\sum_{\{\tilde{\phi}\}}W_{path}[\{\tilde{\phi}\}]}, 
\end{equation}
which satisfies the normalized condition
\begin{equation}
\sum_{\mbox{(all paths)}}\sum_{\{\tilde{\phi}\}}\rho_{path}[\{\tilde{\phi}\}]=1.
\end{equation}
Then the propagator $G_{\sigma}$ in (\ref{epath}) can be reexpressed as
follows 
\begin{equation}\label{egg} 
G_{\sigma}(j,i; E)= \tilde{G}_{\sigma}(j,i; E) \left\langle (-1)^{N^{\downarrow}_{path}}\right\rangle, 
\end{equation} 
where
\begin{equation}\label{eggt}
\tilde{G}_{\sigma}(j,i; E)\equiv - (\sigma)^{j-i}\sum_{\mbox{(all paths)}}\sum_{\{\tilde{\phi}\}}W_{path}[\{\tilde{\phi}\}], 
\end{equation}
and 
\begin{equation}
\left\langle (-1)^{N^{\downarrow}_{path}}\right\rangle 
\equiv \sum_{\mbox{(all paths)}}\sum_{\{\tilde{\phi}\}}\rho_{path}[\{\tilde{\phi}\}] 
\left((-1)^{N_{path}^{\downarrow}}\right).
\end{equation}
Here 
\begin{equation}\label{epstring}
(-1)^{N_{path}^{\downarrow}}\equiv \prod_{s=1}^{K_{ij}}\sigma_{m_s},
\end{equation}
with $N_{path}^{\downarrow}$ denoting the total number of $\downarrow$ spins ``exchanged'' with  the hole as it moves from $i$ to $j$. Notice that $(-1)^{N_{path}^{\downarrow}+ N_{path}^{\uparrow}}\equiv (-1)^{j-i}$ which is 
independent of the
path and thus the system is symmetric about $\uparrow$ and $\downarrow$ 
spins. $\tilde{G}_{\sigma}(j,i; E)$ defined in (\ref{eggt}) may be regarded as 
the single-hole propagator under a new Hamiltonian $\tilde{H}_{t-J}$ obtained by replacing the hopping term $H_t$ in the $t-J$ model with $\tilde{H}_t$, 
whose matrix element is negative-definite  without the extra sign problem 
shown in (\ref{e3}), namely,
\begin{equation}\label{ehtt}
\langle\phi; (m)|\tilde{H}_t|\phi;(n)\rangle=-t.
\end{equation}

One can see from the propagator (\ref{egg}) that a sequence 
of signs $\prod_{s=1}^{K_{ij}}\sigma_{m_s}= (\pm 1)\times (\pm 1) 
\times ... \times (\pm 1)\equiv (-1)^{N^{\downarrow}_{path}}$ is
picked up by the hole from the spin
background. A sign-definite $W_{path}$ or $\rho_{path}$ means that  such a  
phase string cannot be ``repaired'', since there does not exist an other 
source of ``phases''  at low energy to compensate it. In particular, if one 
chooses $i=j$, then all the paths become closed loops on the lattice, and the
gauge-invariant phase $(-1)^{N^{\downarrow}_{path}}$ (which is independent of 
the ways in which one accounts for  the Marshall sign) can be regarded as 
a Berry phase (see Sec. III. A). This Berry phase is incompatible with  a 
quasiparticle picture, in which the whole system should simply get back to the 
original state without picking up a Berry phase each time the quasiparticle returns 
to its original position. 
 Due  to the superposition of such phases from  different paths 
and spin  configurations as shown in (\ref{egg}), it is expected that
the long-distance behavior of the hole will be dramatically modified by the
quantum interference effect of the phase strings. In the following we give a general
proof that the spectral weight which measures the quasiparticle weight of the
injected hole must vanish at the ground-state energy as a direct consequence of
such a phase string effect.

Before going to the next section, we remark that the origin of
this phase string can  be traced back to a highly-nontrivial competition 
between the exchange and hopping processes represented by (\ref{e2}) and 
(\ref{e3}). Recall that each hopping of the hole displaces a spin, leading to a 
spin mismatch. 
Since there are three components for each $SU(2)$ spin which do not commute
with each other, the induced spin-mismatch string has three components in spin
space which must be repaired simultaneously after the hole moves away as 
pointed out in the Introduction.
The phase string effect revealed in (\ref{egg}), however, implies that the 
spin mismatch induced by hopping cannot relax back {\it completely},  and there
is always a residual $U(1)$ phase string left behind, which
is not repairable by low-lying spin fluctuations. This  
subtle phase string effect has been overlooked before, especially  in the 2D case.

\subsection{Phase-string effect: vanishing spectral weight $Z(E_G)$}

First, in momentum space the imaginary part of $G_{\sigma}(k, E)$ can be shown to be 
\begin{equation}\label{eim} 
\mbox{Im} G_{\sigma}(k, E)=-\pi \sum_M Z_k(E_M) \delta(E-E_M),  
\end{equation} 
where the spectral weight $Z_k$ is defined as 
\begin{equation} 
Z_k(E_M)=|\langle\psi_M|c_{k\sigma}|\psi_0\rangle|^2, 
\end{equation} 
with $|\psi_M\rangle$ and $E_M$ denoting the eigenstate and energy of $H_{t-J}$ in  
the one-hole case. 

The corresponding real-space form of (\ref{eim}) is 
\begin{equation}\label{eg"} 
G_{\sigma}''(j,i;E)=-\pi \sum_k e^{-ik\cdot (x_j-x_i)}Z_k(E) \rho(E), 
\end{equation} 
where $\rho(E)=\sum_M\delta(E-E_M)$ is the density of states, and   
$Z_k(E)$ is understood here as being averaged over $M$ at the same energy $E_M=E$. 
If low-lying  
excitations can be classified as quasiparticle-like, one must have a finite
spectral weight at the ground state and its vicinity. Correspondingly, 
$G_{\sigma}''$ should generally be finite when $E\rightarrow E_G$ from $E>E_G$ 
side in two dimensions.\cite{remark} On the other hand, $G_{\sigma}''\equiv 0$ 
at $E<E_G$.

The real part of $G_{\sigma}(k, E)$ 
in the real space can be expressed in terms of $G_{\sigma}''$ by the following
Kramers-Kronig relation:
\begin{equation}\label{ekk}
G'_{\sigma}(j,i; E)=-P\int \frac{dE'}{\pi} \frac{G''(j,i; E')}{E-E'},
\end{equation}
where $P$ denotes taking the principal value of the integral. 
It is straightforward to check that $G'_{\sigma}(j,i; E)$  diverges
logarithmically at $E\rightarrow E_G$ if $G''(j,i; E)$ remains finite at 
$E=E_G^+$:
\begin{equation}
G'_{\sigma}(j,i; E)\sim -\frac{1}{\pi}G''_{\sigma}(j,i;E_G) \ln |E-E_G|.
\end{equation}
On the other hand, by using the spectral expression
\begin{equation}\label{espectral}
G_{\sigma}(j,i; E+ i0^+)=-\int \frac{dE'}{\pi} \frac{G''(j,i; E')}{E-E'+i0^+},
\end{equation}
one finds the analytic continuation of $G_{\sigma}(j,i; E)$ from the upper-half
complex plane to the real axis at $E>E_G$ to be generally well-defined except 
at $E=E_G$. 

Now we discuss the phase string effect. For this purpose, we introduce
the following quantities
\begin{equation}\label{ee}
G_{\sigma}^{e\downarrow}(j,i; E)\equiv -(\sigma)^{j-i}\sum_{\mbox{(all paths)}}\sum_{\{\tilde{\phi}\}}W_{path}[\{\tilde{\phi}\}] \left(\delta_{N_{path}^{\downarrow}, \mbox{even}}\right), 
\end{equation}
with $\delta_{N_{path}^{\downarrow}, \mbox{even}}=1$ if $N_{path}^{\downarrow}=\mbox{even}$
and $\delta_{N_{path}^{\downarrow}, \mbox{even}}=0$ if $N_{path}^{\downarrow}=\mbox{odd}$.
Similarly,
\begin{equation}\label{eo}
G_{\sigma}^{o\downarrow}(j,i; E)\equiv -(\sigma)^{j-i}\sum_{\mbox{(all paths)}}\sum_{\{\tilde{\phi}\}}W_{path}[\{\tilde{\phi}\}] \left(\delta_{N_{path}^{\downarrow}, \mbox{odd}}\right). 
\end{equation}
One may also  define $G_{\sigma}^{e\uparrow}(j,i;E)$ and
$G_{\sigma}^{o\uparrow}(j,i; E)$ in a similar way. Physically, 
$G_{\sigma}^{e,o \uparrow}$ and $G_{\sigma}^{e,o\downarrow}$ measure the 
weights for even or odd number of $\uparrow$ and  $\downarrow$ spins 
encountered by the hole during its propagation from site $i$ to $j$. It is important to note that,  according
to their definition, $G_{\sigma}^{e,o \uparrow}$ and $G_{\sigma}^{e,o\downarrow}$ 
should behave qualitatively  similar in the case of  a symmetric system.   $G_{\sigma}$ in
(\ref{egg}) and $\tilde{G}_{\sigma}$ in (\ref{eggt}) can be then rewritten as 
\begin{equation}\label{eg}
G_{\sigma}(j,i;E)= G_{\sigma}^{e\downarrow}(j,i;E)-G_{\sigma}^{o\downarrow}
(j,i;E),
\end{equation}
and
\begin{equation}\label{egt}
\tilde{G}_{\sigma}(j,i;E)= G_{\sigma}^{e\downarrow}(j,i;E)+G_{\sigma}^{o\downarrow}
(j,i;E).
\end{equation}
Thus $G_{\sigma}^{e\downarrow}(E)$ and $G_{\sigma}^{o\downarrow}(E)$
determine both $G_{\sigma}(E)$ and $\tilde{G}_{\sigma}(E)$, and the phase-string 
effect is simply represented by a minus sign in front of 
$G_{\sigma}^{o\downarrow}(E)$ in (\ref{eg}).

Here a crucial observation is that the ground-state energy $\tilde{E}_G$ of 
$\tilde{H}_{t-J}$ is always lower than the ground-state energy 
$E_G$ of $H_{t-J}$ since, according to the definition in (\ref{ehtt}),
there is no sign problem in  $\tilde{H}_{t-J}$. 
Suppose that the expansions (\ref{ee}) and (\ref{eo}) for $G_{\sigma}^{e\downarrow}(E)$ and $G_{\sigma}^{o\downarrow}(E)$ converge
below some energy $E_0$.
By increasing $E$ the expansions (\ref{ee}) and (\ref{eo}) will eventually  
diverge at $E_0$ with the {\it same} sign because $W_{path}\geq 0$. 
Correspondingly $\tilde{G}_{\sigma}(E)$ also has to diverge at the same energy 
$E_0$ according to (\ref{egt}). It means that $E_0=\tilde{E}_G$ as
$\tilde{G}_{\sigma}(E)$ is analytic at $E<\tilde{E}_G$.  In contrast, 
$G_{\sigma}(E)$ should be still  well-defined at $\tilde{E}_G$  
(note that $E_G>\tilde{E}_G$). Thus the divergent parts in $G_{\sigma}^{e\downarrow}(\tilde{E}_G)$ and 
$G_{\sigma}^{o\downarrow}(\tilde{E}_G)$ have to  cancel out exactly in 
(\ref{eg}). This cancellation is easily understandable, since there is no qualitative 
difference between $G_{\sigma}^{e\downarrow}(E)$ and $G_{\sigma}^{o\downarrow}(E)$. Note that the divergence in (\ref{ee}) and 
(\ref{eo}) is contributed to by all of those 
paths connecting the fixed $i$ and $j$ whose lengths approach infinity.
In this limit, the effects of the even or
odd total number of $\downarrow$ spins on the hole's  path become indistinguishable.

But we are mainly interested in the behavior of $G_{\sigma}(E)$ near $E\sim E_G$. 
According to the previous discussion, for a finite  spectral weight $Z_k(E_G)$
the real part of $G_{\sigma}(E)$ has to diverge at $E=E_G$. On the other hand, 
the analytic continuation of $\tilde{G}_{\sigma}(E)$ to $E_G+i0^+$ should
remain well-defined in terms of the spectral expression similar to 
(\ref{espectral}). In other words,  if $Z_k(E_G)\neq 0$, one should find that
$G_{\sigma}^{e\downarrow}(E)$ and  $G_{\sigma}^{o\downarrow}(E)$ 
(after an analytic continuation across the upper half plane to the real axis 
at $E>\tilde{E}_G$) have to  diverge again at $E=E_G$ in the following way:
\begin{equation}\label{ediv1}
G_{\sigma}^{e\downarrow}(E_G)-G_{\sigma}^{o\downarrow}(E_G)\rightarrow \infty,
\end{equation}
whereas 
\begin{equation}\label{ediv2}
G_{\sigma}^{e\downarrow}(E_G)+G_{\sigma}^{o\downarrow}(E_G)= {\mbox {finite}}.
\end{equation}
However, this would mean that $G_{\sigma}^{e\downarrow}(E_G)$ and 
$G_{\sigma}^{o\downarrow}(E_G)$ have to diverge with {\it opposite} signs,  which is contrary to the 
intuitive observations (recall that both of them have 
the same sign at $E<\tilde{E}_G$ as defined in (\ref{ee}) and (\ref{eo}) and 
diverge with the same sign at $\tilde{E}_G$ as discussed earlier on). Such behavior also means 
a violation of spin symmetries of the system. Let us consider  $G^{e\uparrow}_{\sigma}$ and 
$G^{o\uparrow}_{\sigma}$, characterizing the contributions from $\uparrow$ 
spins, whose definitions are similar to (\ref{ee}) and (\ref{eo}). Suppose
that $i$ and $j$ belong to  different sublattice sites. A simple counting then 
shows that $N^{\uparrow}_{path}+N^{\downarrow}_{path}=$ odd integer, and one 
finds that $G^{e\uparrow}_{\sigma}(E)=G^{o\downarrow}_{\sigma}(E)$ and $
G^{o\uparrow}_{\sigma}(E)=G^{e\downarrow}_{\sigma}(E)$.
In terms of  (\ref{ediv1}) and (\ref{ediv2}), then,  they should diverge with {\it opposite} signs too, namely, $G^{e\downarrow}_{\sigma}(E)\rightarrow \infty$ and $G^{e\uparrow}_{\sigma}(E)\rightarrow
-\infty$ at $E\rightarrow E_G$,  However, 
according to their definitions, this indicates a violation of 
spin symmetries at $E\rightarrow E_G$, with  contributions from $\uparrow$ and $\downarrow$ spins 
behaving drastically different in contrast to their symmetric definition at $E \leq\tilde{E}_G$. 
Therefore, one has to conclude that $G'_{\sigma}(j,i; E)$ cannot diverge at
$E_G$ due to the phase string effect, which indicates  that the
spectral weight $Z(E_G)$ has to vanish at low-dimensions where the density of
states $\rho(E)\neq 0$ at $E=E_G$. 

The way that the phase-string effect leads to the 
vanishing of $Z(E_G)$ can be also
intuitively understood in an another way.\cite{sheng} Notice that the phase string factor 
defined in (\ref{epstring}) is quite singular as it changes  sign each time when
the total number $N_{path}^{\downarrow}$ increases or decreases by one, no matter
how long the path is. But it would become 
meaningless to distinguish even and odd number of $\downarrow$ spins
encountered by the hole when the path is infinitely long. Consequently, the
average $\langle(-1)^{N^{\downarrow}_{path}}\rangle$ has to vanish at $|i-j|
\rightarrow \infty$, (since $(-1)^{N^{\downarrow}_{path}}= +1 $ for even 
$N^{\downarrow}_{path}$ and  $(-1)^{N^{\downarrow}_{path}}= -1 $ for odd 
$N^{\downarrow}_{path}$ will have  equal probability at this limit). Due to such phase-string 
frustration, the propagator (\ref{eg}) will always decay {\it faster } than a 
regular quasiparticle-like one (i.e., $\tilde{G}$) at large distance, and in particular, it  has to keep decaying
even  at the ground-state energy $E_G$  
which  then requires a vanishing $Z(E)$ at $E=E_G$ as  can be  shown in terms of (\ref{eg"}).  

$Z(E_G)=0$ means that there is no direct overlap between the ``bare'' hole 
state $c_{i\sigma}|\psi_0\rangle$ and the true  ground state. Thus the behavior of a hole injected 
into the undoped  ground state is indeed dramatically modified by 
the  phase string effect, as compared to its quasiparticle-like bare-hole state. It implies the failure 
of a conventional perturbative approach to this problem which generally requires a zeroth-order
overlap of  the bare state with the true ground state. We would like to note that even though exact
diagonalization calculations on small lattices\cite{dagotto,num} have indicated a 
quasiparticle peak at the energy bottom of the spectral function, when the 
lattice size goes to infinity,  it is hard 
to tell from the small-size numerics whether such a quasiparticle peak would still stay at the
lower end of the spectra or there could be some weight (e.g., a tail) emerging 
below the peak which vanishes at the ground-state energy (such that 
$Z(E_G)=0$). The present analysis suggests that the large-scale effect is 
really important in this system due to the phase string effect. Therefore,
finite-size numerical calculations as well as various analytical approaches
should be under scrutiny with regard to the long-wavelength, low-energy 
properties.

The conditions (\ref{e2}) and  (\ref{e3}) are crucially responsible for producing the non-repairable phase-string effect in the above discussion. These are the 
intrinsic properties of the $t-J$ Hamiltonian itself. On the other hand,  the condition that $|\psi_0\rangle$ is {\it  the ground state of the undoped  antiferromagnet} actually does not play a crucial role in the demonstration 
of $Z(E_G)=0$. In other words, the whole argument about $Z(E_G)=0$  should still remain robust even 
when $|\psi_0\rangle$ is replaced by a general ground state at finite 
doping. Of course, at finite doping some additional phase effect due to
the fermionic statistics among holes will appear in the matrix (\ref{e3}), but 
such a sign  problem should not invalidate the phase-string effect at least at small doping concentrations.

\section{PHASE STRING EFFECT AT FINITE DOPING}

The non-repairable phase string effect exhibited in the single-hole propagator
({\ref{egg}) reveals a remarkable competition 
between the superexchange and hopping processes in the $t-J$ model. This 
effect leads to the breakdown of conventional perturbative methods as discussed
in the one-hole case. We have also pointed out that such a phase
string effect is generally present even at finite doping. In this section, we
will introduce a useful mathematical formalism to 
describe this effect in the presence of a finite amount of holes.

\subsection{Phase string effect and the Berry phase}

The phase string is defined  as a product of a  sequence of signs
\begin{equation}\label{ephase}
(\pm 1)\times (\pm 1)\times ...(\pm 1)
\end{equation}
where $(\pm 1)\equiv \sigma_m $ is decided by  the {\it instant} spin
$\sigma_m$ at a site $m$ at the moment when the hole hops 
to that site. So the phase string depends on both the hole path as well as 
the instant spin configurations. As shown in the propagator (\ref{egg}), such a 
phase string is always picked up by the hopping of the hole from
the quantum spin background. In particular, if the hole moves through a 
closed-path $C$ on the lattice back to its original position, it
will get a phase $(-1)^{N^{\downarrow}_C}$,
where $N^{\downarrow}_C$ is the total number of $\downarrow$ spins 
``encountered'' by the hole on the closed-path $C$. (It is noted that $(-1)^{N^{\downarrow}_C}=(-1)^{N^{\uparrow}_C}$ as a closed path $C$ always 
involves an even number of lattice sites. So there is no 
symmetry violation even though we will focus on $(-1)^{N^{\downarrow}_C}$ 
below.)  If one lets the hole move slowly enough on the path $C$ such that the 
spin-displacement created by its motion is able to relax back, then after the
hole returns to its original position, the whole system will restore back to
the original one except for 
an additional phase $(-1)^{N^{\downarrow}_C}$. Thus, the closed-path phase 
string $(-1)^{N^{\downarrow}_C}$ may be 
regarded as a Berry phase.\cite{berry}
Of course, here $(-1)^{N^{\downarrow}_C}$ is not simply a geometric phase which is only path-dependent, but also depends on the  
spin configurations along its path. 

To keep track of such a Berry phase in the ground-state wavefunction, we may 
introduce the following quantity
\begin{equation}\label{eberry}
e^{i\Theta}=\exp\left\{-i\sum_{i,l}\mbox{Im ln $ (z_i^{(h)}- z_l^{(b\downarrow)})$}\right\},
\end{equation}
where $z\equiv x+iy$ with superscripts $(h)$ and $(b\downarrow)$ refer
to hole and $\downarrow$ spin, respectively, and the subscripts $i$ and $l$ denoting their lattice sites. 
Here the definition is not restricted to the one hole case, and 
$z_i^{(h)}\neq z_l^{(b\downarrow)}$ is guaranteed by the no-double-occupancy
constraint. Let us consider the evolution of $e^{i\Theta}$ under a closed-loop 
motion for a given hole on the lattice. Recall that at each step of hole 
hopping, the spin originally located at the new hole site has to be transferred back to the original hole site. If it is a $\downarrow$ spin, then 
$\mbox{Im ln $ (z_i^{(h)}- z_l^{(b\downarrow)})$}$ in (\ref{eberry}) will give 
rise to a phase-shift $\pm \pi$ due to such an ``exchange'' and thus a $(-1)$ 
factor in $e^{i\Theta}$. After the hole returns to its original position 
through the closed-path $C$ and all the displacement of spins on the path
is restored (which can be realized through spin flips as discussed before),
one finds
\begin{equation}\label{eberry1}
e^{i\Theta}\rightarrow (-1)^{N^{\downarrow}_C} \times e^{i\Theta}.
\end{equation}
(Each $\downarrow$ spin inside the closed-path $C$ contributes 
a phase $e^{\pm i 2\pi}=1$.) Thus, the phase factor $e^{i\Theta}$ 
reproduces the afore-mentioned Berry phase due to the phase string effect.

Therefore, it is natural to incorporate the phase factor $e^{i\Theta}$ explicitly 
in the wavefunction to track the Berry phase, or define the following new spin-hole basis
\begin{equation}\label{ebasis}
|\bar{\phi}\rangle=e^{i\hat{\Theta}}|\phi\rangle,
\end{equation}
where $e^{i\hat{\Theta}}$ is the operator form of $e^{i\Theta}$ in 
(\ref{eberry}):
\begin{equation}\label{ephase2} 
e^{i\hat{\Theta}}\equiv  e^{-i\sum_{i,l }n_i^h\left[\theta_i(l)\right] 
n_{l\downarrow}^b},
\end{equation}
in which $n^h_i$ and $n_{l\downarrow}^b$ are defined as the hole and 
$\downarrow $-spin occupation number operators, respectively, with  
$\theta_i(l)$ defined by
\begin{equation}\label{ebph2}
\theta_i(l)=\mbox{Im ln $(z_i-z_l)$}.
\end{equation}
By using the no-double-occupancy constraint, one may also rewrite 
$n_{l\downarrow }^b$ as 
$n_{l\downarrow }^b=\frac 1 2 \left[1- n^h_l-\sum_{\sigma}\sigma n_{l\sigma}^b\right]$
and thus express $e^{i\hat{\Theta}}$ in a symmetric form with regard to 
$\uparrow$ and $\downarrow$ spins:
\begin{equation}\label{ephase3} 
e^{i\hat{\Theta}}\equiv  e^{-i\frac 1 2 \sum_{i,l}n_i^h\theta_i(l) 
\left[1- n^h_l-\sum_{\sigma}\sigma n_{l\sigma}^b\right]}.
\end{equation}
According to previous discussions, this new basis 
should be more appropriate for expanding the true ground state 
$|\psi_G\rangle$ as well as the low-lying states because the hidden Berry phase 
due to the phase string effect is explicitly tracked. 
In other words, the wavefunction $\bar {\chi}_{\phi}$ defined in  
$|\psi_G\rangle=\sum_{\phi}\bar {\chi}_{\phi}|\bar{\phi}\rangle$ should become more or less
``conventional'' as the singular phase string effect is now sorted out into $|\bar{\phi}\rangle$. 
Correspondingly the Hamiltonian in this new representation is expected to be 
perturbatively treatable as the phase string effect is ``gauged away'' by the
singular gauge transformation in (\ref{ebasis}). In the following section, we reformulate the $t-J$ model in this new representation for an arbitrary number  of holes.   

\subsection{``Phase-string'' representation of the $t-J$ model} 

We start by generalizing the spin-hole basis (\ref{esh}) to an arbitrary hole 
number $N_h$ in the Schwinger-boson, slave-fermion representation:
\begin{equation}\label{esh1}
|\phi\rangle=(-1)^{N^{\downarrow}_A} \left\{b^{\dagger}_{i_1\uparrow}...
b^{\dagger}_{i_{M}\uparrow}\right\}\left\{b^{\dagger}_{i_{M+1}\downarrow}...
b^{\dagger}_{i_{N_e}\downarrow}\right\}\left\{f^{\dagger}_{l_1}...f^{\dagger}_{
l_{N_h}}\right\}|0\rangle,
\end{equation}
where $N_e$ is the total electron (spin) number, and the fermionic ``holon'' 
creation operator $f_i^{\dagger}$ and the bosonic 
``spinon'' annihilation operator $b_{i\sigma}$ commute with each other, 
satisfying the no double occupancy constraint
\begin{equation}\label{econstraint}
f^{\dagger}_if_i+\sum_{\sigma}b^{\dagger}_{i\sigma}b_{i\sigma}=1.
\end{equation}
The electron operator is written in this formalism as $c_{i\sigma}= f_i^{
\dagger}b_{i\sigma}$. Now if we redefine the spinon operator as $b_{i\sigma}
\rightarrow (-\sigma)^i b_{i\sigma}$ such that  
\begin{equation}\label{esw}
c_{i\sigma}= f_i^{\dagger}b_{i\sigma}(-\sigma)^i,
\end{equation}
then the Marshall sign in (\ref{esh1}) can be  absorbed into the spinon creation
operators:
\begin{equation}\label{esh2}
|\phi\rangle=(-1)^{N_A}\left\{b^{\dagger}_{i_1\uparrow}...b^{\dagger}_{i_{M}\uparrow}
\right\}\left\{b^{\dagger}_{i_{M+1}\downarrow}...b^{\dagger}_{i_{N_e}\downarrow}\right\}\left\{f^{\dagger}_{l_1}...f^{\dagger}_{l_{N_h}}\right\}|0\rangle,
\end{equation}
with  $(-1)^{N_A}$ being a trivial phase factor left for a later convenience (here $N_A$ denotes the total spin number on the A sublattice site).  In terms of (\ref{esw}), the superexchange term
(\ref{e01}) in the $t-J$ model can be rewritten in the following form after
using the no-double-occupancy constraint:
\begin{equation}
H_J=-\frac J 2 \sum_{\langle ij\rangle \sigma, \sigma'}
b_{i\sigma}^{\dagger}b_{j-\sigma}^{\dagger}b_{j-\sigma'}b_{i\sigma'}.
\end{equation}
It is easy to check that the matrix element $\langle \phi'|H_J|\phi\rangle \leq 0$
for the basis defined in (\ref{esh2}). Namely, the Marshall sign rule is explicitly
built-in here. The hopping term $H_t$ in (\ref{ehop}) becomes
\begin{equation}\label{ehop2}
H_t=-t\sum_{\langle ij\rangle \sigma }\left(\sigma\right)f^{\dagger}_if_j 
b_{j\sigma}^{\dagger}b_{i\sigma}
 + H.c.   
\end{equation}
where the spin index $\sigma$ describing  the sign source generating the phase 
string in (\ref{ephase}) is explicitly shown. Besides the sign-source due to $\sigma$, 
the fermionic statistics of $f_i$ in (\ref{ehop2})
at many-hole cases will also contribute a sign for each exchange of 
two fermions. 

Now we introduce the new spin-hole basis (\ref{ebasis}) and (\ref{ephase3}).
The phase-shift factor $e^{i\hat{\Theta}}$ in (\ref{ephase3}) can be regarded as a 
unitary transformation and any operator $\hat{O}$ should be expressed in the new
representation by a canonical transformation $\hat{O}\rightarrow 
e^{i\Theta}\hat{O}e^{-i{\hat{\Theta}}}$. Then, the hopping term $H_t$ and the
 superexchange term $H_J$ of the $t-J$ model in the slave-fermion 
representation can be expressed under this transformation as follows
\begin{equation}\label{et}
H_t=-t\sum_{\langle ij\rangle \sigma}\left(e^{iA_{ij}^f}\right)h^{\dagger}_ih_j
\left(e^{i\sigma A_{ji}^h}\right)b_{j\sigma}^{\dagger}b_{i\sigma} + H.c.   
\end{equation}
and 
\begin{equation}\label{ej}
H_J=-\frac J 2 \sum_{\langle ij\rangle \sigma \sigma'}
\left(e^{i\sigma
A_{ij}^h}\right)b_{i\sigma}^{\dagger}b^{\dagger}_{j-\sigma}\left(e^{i\sigma' 
A_{ji}^h}\right)b_{j-\sigma'}b_{i\sigma'}.
\end{equation}
in which gauge phases $A_{ij}^f$ and $A_{ij}^h$ are defined by
\begin{equation}\label{eaf}
A_{ij}^f=\frac 1 2 \sum_{l\neq i, j}\left[\theta_i(l)-\theta_j(l)
\right]\left(\sum_{\sigma}\sigma n_{l\sigma}^b-1\right),
\end{equation}
and
\begin{equation}\label{eah}
A_{ij}^h=\frac 1 2 \sum_{l\neq i, j}\left[\theta_i(l)-\theta_j(l)\right]n_l^h.
\end{equation}
Here $h_i$ is defined by
\begin{equation}\label{ejw}
h_i^{\dagger}=f_i^{\dagger}\left[e^{-i\sum_{l\neq i}\theta_i(l) n_l^h}\right].
\end{equation}
Eq.(\ref{ejw}) actually represents a Jordan-Wigner transformation\cite{jw} which 
changes the fermionic statistics of $f_i^{\dagger}$ into the bosonic operator 
$h_i^{\dagger}$. So both $h_i$ and $b_{i\sigma}$ now are bosonic operators in 
this new representation. Note that the sign factor 
$\sigma$ appearing in the slave-fermion formalism of $H_t$ in (\ref{ehop2}) 
no longer shows up in (\ref{et}),  which means that the phase string is 
indeed ``gauged away''.  Nevertheless, one gets nonlocal gauge fields
$A_{ij}^f$ and $A_{ij}^h$ in the new representation. In the one dimensional case, 
$A_{ij}^f=A_{ij}^h=0$ (see the section IV. A), but they are nontrivial in 
two dimensions. For example, for a counter-clockwise-direction closed-path $C$
on 2D lattice, one finds   
\begin{equation}\label{eflux1}
\sum_C A_{ij}^f= \pi\sum_{l\in C}\left(\sum_{\sigma}\sigma n^b_{l\sigma}-1\right)+ \mbox{$\sum_C^f$},
\end{equation}
and
\begin{equation}\label{eflux2}
\sum_C A_{ij}^h=\pi\sum_{l\in C} n_l^h+\mbox{$\sum_C^h$}, 
\end{equation}
where the notation $l\in C$ on the right-hand side means that the summations
are over the sites inside the path $C$,  while $\sum_C^f$ and  $\sum_C^h$ denote the contributions from 
the sites right on the path $C$,  the latter being  different from those inside the path $C$  by a numerical 
factor $\frac 1 2$ or $\frac 1 4$ depending on whether they are at the corner or along the edge of 
the closed path $C$.  Nonzero  (\ref{eflux1}) and (\ref{eflux2})  show that $A_{ij}^f$ and $A_{ij}^h$,  which cannot be gauged away in 2D,  describe vortices (quantized flux tubes) centered on spinons (in fact, 
$A_{ij}^f$ also includes an additional lattice $\pi$-flux-per-plaquette as represented by the
second term in the first summation on the right hand side of (\ref{eflux1})) and holons, respectively.  
Physically, this suggests the existence of nonlocal correlations between the 
charge and spin degrees of freedom in 2D. For instance, in $H_J$ (\ref{ej})
spins can always feel the effect of holes nonlocally through the gauge field
$A_{ij}^h$. It is a direct consequence of the phase string effect, which depends
on both the hole path as well as the instant spin configurations on the path.

Furthermore, the slave-fermion 
decomposition of electron operator in (\ref{esw}) is transformed in terms of 
$c_{i\sigma}\rightarrow e^{i{\hat{\Theta}}}c_{i\sigma}e^{-i{\hat{\Theta}}}$ as 
follows\cite{remark3}
\begin{equation}\label{emutual}
c_{i\sigma}=\tilde{h}_i^{\dagger}\tilde{b}_{i\sigma}(-\sigma)^i,
\end{equation}
in which
\begin{equation}\label{espinon}
\tilde{b}_{i\sigma}\equiv b_{i\sigma}\left[e^{-i\frac {\sigma} 2 \sum_{l\neq i} 
\theta_i(l) n_l^h}\right], 
\end{equation} 
and
\begin{equation}\label{eholon}
\tilde{h}_i^{\dagger}\equiv h_i^{\dagger}\left[e^{i\frac 1 2 \sum_{l\neq i} 
\theta_i(l)\left(\sum_{\alpha}\alpha n_{l\alpha}^b -1\right)} \right]. 
\end{equation} 

This new decomposition form is quite non-conventional as it involves nonlocal
Jordan-Wigner type phase factors in the ``spinon'' annihilation operator 
$\tilde{b}_{i\sigma}$ and  ``holon'' 
creation operator $ \tilde{h}^{\dagger}_i$.  One can easily check that  
$\tilde{b}_{l\sigma}^{\dagger}$ and $ \tilde{h_j}^{\dagger}$ satisfy the
following mutual statistics
\begin{equation}\label{ecomm}
\tilde{b}_{l\sigma}^{\dagger}\tilde{h}^{\dagger}_j = \left(\pm i\sigma \right)
\tilde{h}^{\dagger}_j \tilde{b}_{l\sigma}^{\dagger},
\end{equation}
etc., for $l\neq j$.  Here signs $\pm$ denote  two different 
ways (clock-wise and counter-clock-wise) by which the spinon and holon 
operators are exchanged. Because of the phase $\pm i$, spinons and holons 
defined here obey ``semion''-like mutual statistics, and $\uparrow$ 
and  $\downarrow$ spinons show opposite signs in the commutation relation 
(\ref{ecomm}). Thus the present ``phase-string'' representation may also be
properly regarded as  a ``mutual-statistics'' decomposition. The physical origin of 
the mutual statistics may be understood based on the phase string effect. 
As defined in (\ref{ephase}),  a phase string factor changes when and only when 
hopping, i.e., {\it an exchange of a spin and a hole}, takes place.  
Thus, it can be described as a {\it counting} problem. One may keep track of 
such a phase string  exactly by letting the hole and spin 
satisfy a {\it mutual statistics} relation, such that an exchange of a hole with a spin 
$\sigma_m$ should produce an extra phase depending on $\sigma_m$, as shown 
in (\ref{ephase}). The role of the phase string effect may be then regarded as 
to  simply induce a  mutual statistics between the spin and charge degrees of 
freedom. In this way, the phase string itself may be ``gauged away'' from the 
Hamiltonian, but at the price of dealing with a mutual statistics problem. 
Furthermore, (\ref{espinon}) and (\ref{eholon}) can be understood 
as composite-particle expressions\cite{wilczek} for the
 mutual-statistics spinon and holon ($\tilde{b}_{i\sigma}$ and $\tilde{h}_i$), in terms of conventional bosons ($b_{\sigma}$ and $h_i$) carrying flux tubes. 
In other words,  we still work in a conventional bosonic
representation of spinon and holon where the mutual statistics effect is 
transformed to an interaction problem, which is  similar to a fractional-statistics 
system.\cite{wilczek,laughlin}  This may be seen 
from the corresponding Hamiltonians of (\ref{et}) and (\ref{ej}) in the new  
representation,  with the gauge fields $A_{ij}^h$ and $A_{ij}^f$ representing 
mutual statistics effect. 
    
Therefore, after explicitly sorting out the phase string effect, the $t-J$
Hamiltonian is reformulated in (\ref{et}) and (\ref{ej}), where the original
singular phase string effect is ``gauged away'' in 1D, while its residual 
effect is  represented by nonlocal
gauge fields $A_{ij}^h$ and $A^f_{ij}$ in 2D. As far as long-distant physics
is concerned, fluctuations of $A_{ij}^h$ and $A^f_{ij}$ are presumably
small for  $(ij)\in $ the nearest-neighbors, as compared to the original singular phase
string effect. Physically, this is due to the fact that only a quantum superposition effect
of all the phase strings from different paths contributes to the energy, which behaves
relatively ``mildly'' as discribed by  $A_{ij}^h$ and $A^f_{ij}$.  Then a generalized 
local mean-field type approximation (examples are to be given in the next section) may 
become  applicable to this  new formalism. On the other hand, the singular part of 
the phase string effect is now kept in the electron $c$-operator expression (\ref{emutual}). 
Thus, in drastically contrast to the conventional picture, 
the ``test'' particles created by physical operators (as combinations of the electron  $c$-operators) 
on the ground state will not simply ``decay'' into 
{\it internal} elementary charge and spin excitations (known as holons and
spinons here). The  nonlocal phase in the decomposition (\ref{emutual}) will
change their nature in  a fundamental  way, as will be illustrated by  the 1D example
given below.   The relation of ``test'' particles to  internal elementary excitations in Fermi-liquid 
systems is completely  changed here by the phase string effect. 
Thus, knowing internal holon and spinon excitations no longer means that various 
electron correlation functions can be automatically determined, to leading order,  from a simple 
convolution of their propagators.

\section{EXAMPLES}

In the last section, a mathematical formalism was established for a general doping concentration,
in which the phase string effect was explicitly tracked. In
the following we apply this formalism to some examples in both 1D and
2D which reveal highly-nontrivial consequences of the  phase string 
effect in the low-energy, long-wavelength regime. 
 
\subsection{1D example: asymptotic correlation functions}

It is well known that the Luttinger-liquid behavior exhibited in this 1D
system is difficult to describe by conventional many-body theories. The 
success of the bosonization approach\cite{bosonization,b1,ra,b2} to this problem relies heavily on the
Bethe-ansatz solution\cite{lieb} of the model. An alternative path-integral approach\cite{weng1}
{\it without using the Bethe-ansatz solution} can also provide a systematic 
understanding of the Luttinger-liquid behavior at $J\ll t$. In this latter
approach, some $U(1)$ nonlocal phase is found to play the key role.
This nonlocal  $U(1)$ phase,  which originated from the coupling of doped holes with 
the $SU(2)$ spins,  has been also shown to be related\cite{weng3} to the Marshall sign hidden in the 
spin background, and is thus directly connected to the phase string effect 
discussed in the present paper.

In this section, we use this 1D system as a nontrivial example and 
outline how the phase string effect can straightforwardly lead to the 
correct leading-order Luttinger-liquid behavior of correlation functions, 
without involving complicate mathematical description usually associated with 
this 1D problem. 

Let us start with the decomposition (\ref{emutual}). 
In the 1D case, $\theta_i(l)$ defined in (\ref{ebph2}) is reduced to
\begin{equation}\label{eth1d}
\theta_i(l)=\left\{\begin{array}{ll} \pm \pi, & \mbox{  if $i< l$},\\
0,  & \mbox{if $i> l$}.
\end{array}\right.
\end{equation}
According to (\ref{espinon}) and (\ref{eholon}), the decomposition 
(\ref{emutual}) can be then written as
\begin{equation}\label{e1d}
c_{i\sigma}=h^{\dagger}_ib_{i\sigma} e^{\pm i\left[\sigma \Theta_i^h+ 
\Theta_i^b\right]},
\end{equation}
where
\begin{equation}\label{ephh}
\Theta^h_i=\frac{\pi}{2}\sum_{l>i}\left(1-n_l^h\right) ,
\end{equation}
and
\begin{equation}\label{ephb}
\Theta^b_i=\frac{\pi}{2} \sum_{l>i, \alpha} \alpha n_{l\alpha}^b,
\end{equation}
(Note that in (\ref{e1d}) a phase factor $(-\sigma)^ie^{\mp i\pi (1+\sigma)/2
\sum_{l>i}}$ is omitted  which can be easily shown to be equal to $1$ for a bipartite lattice.)

In terms of (\ref{eth1d}), it
is straightforward to show that $A_{ij}^f=A_{ij}^h=0$,  and thus (\ref{et}) and
(\ref{ej}) reduce to 
\begin{equation}\label{et1d}
H_t=-t\sum_{\langle ij\rangle \sigma}h^{\dagger}_ih_j b_{j\sigma}^{\dagger}b_{i\sigma} + H.c.   
\end{equation}
and 
\begin{equation}\label{ej1d}
H_J=-\frac J 2 \sum_{\langle ij\rangle \sigma, \sigma'}
b_{i\sigma}^{\dagger}b^{\dagger}_{j-\sigma}b_{j-\sigma'}b_{i\sigma'}.
\end{equation}
Therefore, there is no sign problem (phase frustration) present in the
Hamiltonian since both holon and spinon are bosonic here. {\it All of the 
important phases are exactly tracked by the phase
$e^{\pm i\left[\sigma \Theta_i^h + \Theta_i^b\right]}$ in the decomposition
(\ref{e1d}) which is the explicit expression for the nonlocal phase string in 
the 1D many-hole case.} Without the presence of  the  nonlocal phase effect, a
conventional mean-field type of approximation may become
applicable to the Hamiltonians in the new 
representation. In the 
following we outline a simple mean-field theory for (\ref{et1d}) and 
(\ref{ej1d}).

One may rewrite (\ref{ej1d}) in the following form
\begin{equation}\label{ej1dd}
H_J=-\frac J 2 \sum_{\langle ij\rangle \sigma, \sigma'}
b_{i\sigma}^{\dagger}b_{j\sigma}b^{\dagger}_{j\sigma'}b_{i\sigma'},
\end{equation}
after using $b^{\dagger}_{j-\sigma}b_{j-\sigma}=b_{j\sigma}b^{\dagger}_{j\sigma}$
at an occupied site $j$ where a hard-core condition of $(b^{\dagger}_{j\sigma})^2
=0$ is employed. Then by introducing the mean-fields
$\chi_0=\frac 1 2 \sum_{\sigma}\langle 
b_{j\sigma}^{\dagger}b_{i\sigma}\rangle$ and $H_0=\langle h_i^{\dagger}h_j
\rangle $ with $i$ and $j$ being the nearest-neighboring sites, a mean-field
version of the $t-J$ Hamiltonian $H_{eff}=H_h+H_s$ can be obtained with
\begin{equation}\label{eh}
H_h=-t_h\sum_{\langle ij \rangle}h^{\dagger}_i h_j + H.c.,
\end{equation}
and 
\begin{equation}\label{es}
H_s=-J_s\sum_{\langle ij\rangle \sigma} b_{i\sigma}^{\dagger}b_{j\sigma}+
H.c.,
\end{equation}
with $t_h=2t\chi_0$ and $J_s=J\chi_0+tH_0$. Such a mean-field solution has
been derived before (Ref.\onlinecite{weng2}). A more accurate mean-field
version may be obtained\cite{weng1} with the same structure as in (\ref{eh}) and 
(\ref{es}), but the summation $\sum$ in $H_s$ now should be understood as over a 
``squeezed spin  chain'' defined by removing hole sites away at any given 
instant at $J\ll t$ limit. In this way the ``hard-core'' constraint between 
the holon and spinon can be automatically satisfied.\cite{weng1} 

$H_h$ and $H_s$ are the tight-binding Hamiltonians for hard-core bosons which can be 
easily diagonalized. For example, one may introduce the Jordan-Wigner transformation
\begin{equation}
h_i=f_i e^{\mp i\pi\sum_{l>i} n_l^h},
\end{equation}
where $f_i$ is a fermionic operator, and correspondingly $H_h$ becomes
\begin{equation}
H_h=-t_h\sum_{\langle ij \rangle}f^{\dagger}_i f_j + H.c.,
\end{equation}
which describes the tight-binding model of free fermions and can be diagonalized in the
momentum space. The solution for $H_s$ is similarly known after using the
Jordan-Wigner transformation $b_{i\sigma}=f_{i\sigma} e^{\mp i \pi 
\sum_{l>i}n_{l\sigma}^b}$. Thus, in principle
one can use the decomposition (\ref{e1d}) to determine various correlation
functions.

Let us consider the single-electron Green's function
\begin{eqnarray}\label{egreen}
\langle c^{\dagger}_{j\uparrow}(t)c_{i\uparrow}(0)\rangle &= & 
\left\langle h_j(t)b^{\dagger}_{j\uparrow}(t) e^{\mp i\left[\Theta_j^h(t)+\Theta_j^b(t)\right]}e^{\pm i\left[\Theta_i^h(0)+
\Theta_i^b(0)\right]}b_{i\uparrow}(0)h_i^{\dagger}(0)\right\rangle \nonumber\\
& = & \left\langle h_j(t) e^{\mp i\Theta_j^h(t)}e^{\pm i\Theta_i^h(0)
}h_i^{\dagger}(0)\right\rangle \left\langle b^{\dagger}_{j\uparrow}(t) e^{\mp i
\Theta_j^b(t)}e^{\pm i
\Theta_i^b(0)}b_{i\uparrow}(0)\right\rangle 
\end{eqnarray}
where a spin-charge separation condition in the ground state of 
$H_{eff}=H_h+H_s$ is 
used. Since both $H_h$ and $H_s$ can be expressed in terms of free-fermion solutions, one may 
use the bosonization\cite{haldane} to describe the quantities involved on the right-hand-side
of (\ref{egreen}) in the long-distance and long-time limit. 
For example, to  
leading order, one finds\cite{weng1} 
\begin{eqnarray}
\left\langle h_j(t) e^{\mp i\Theta_j^h(t)}e^{\pm i\Theta_i^h(0)
}h_i^{\dagger}(0)\right\rangle & =& \left\langle f_j(t) e^{\mp i\frac {\pi}{2}\sum_{l>j}
(1+n_l^h(t))}e^{\pm i\frac{\pi}{2}\sum_{l>i}(1+n_l^h(0))
}f_i^{\dagger}(0)\right\rangle \nonumber\\
&\propto & \frac {e^{\pm ik_fx}} {(x\pm v_h t)^{\frac 1 2}}\left\langle e^{\mp i\frac {\pi}{2}
\sum_{l>j}:n_l^h(t):}e^{\pm i\frac{\pi}{2}\sum_{l>i}:n_l^h(0):
}\right\rangle\nonumber\\
& \propto & \frac {e^{\pm ik_fx}} {(x\pm v_h t)^{\frac 1 2}(x^2-v_h^2t^2)^
{\frac{1}{16}}},
\end{eqnarray}
where $x\equiv x_j-x_i$, $v_h=2t_ha \sin (\pi \delta)$, and $k_f$ is the 
electron Fermi-momentum defined by $k_f=\frac {\pi}{2a}(1-\delta)$. In 
obtaining the last line, the following expression is used:
\begin{eqnarray}\label{egph2}
\left\langle e^{\mp i\frac {\pi}{2}\sum_{l>j}:n_l^h(t):}e^{\pm i\frac{\pi}{2}\sum_{l>i}:n_l^h(0):
} \right\rangle 
\propto \frac{1}{ (x^2-v_h^2t^2)^{\frac{1}{16}}},
\end{eqnarray}
where the normal ordering $:n^h_l:$ is defined as $n^h_l-\langle n^h_l\rangle$. 
The spinon part can be similarly evaluated based on the bosonization
technique: 
\begin{eqnarray}
\left\langle b^{\dagger}_{j\uparrow}(t) e^{\mp i\Theta_j^b(t)}e^{\pm i
\Theta_i^b(0)}b_{i\uparrow}(0)\right\rangle & =&
\left\langle f^{\dagger}_{j\uparrow}(t) e^{\pm i\frac{\pi}{2} \sum_{l>j,\alpha}n_{l\alpha}^b(t)}
e^{\mp i\frac{\pi}{2}
\sum_{l>i,\alpha}n^b_{l\alpha}(0)}f_{i\uparrow}(0)\right\rangle \nonumber\\
&\propto & \frac {1}{(x\pm v_s t)^{\frac 1 2}}\left\langle e^{\pm i\frac{\pi}{2} 
\sum_{l>j,\alpha}:n_{l\alpha}^b(t):}e^{\mp i\frac{\pi}{2}
\sum_{l>i,\alpha}:n^b_{l\alpha}(0):}\right\rangle\nonumber\\
&\propto & \frac {1}{(x\pm v_s t)^{\frac 1 2}},  
\end{eqnarray}
where $v_s=2J_s a/(1-\delta)$, and by using the constraint 
$\sum_{\sigma} n_{l\sigma}^b=1$ on the ``squeezed spin chain'', one has 
\begin{equation}
\left\langle e^{\pm i\frac{\pi}{2} \sum_{l>j,\alpha}:n_{l\alpha}^b(t):}e^{\mp i\frac{\pi}{2}
\sum_{l>i,\alpha}:n^b_{l\alpha}(0):}\right\rangle\sim 1.
\end{equation}
Therefore,   the single-electron Green's function
has the following leading behavior
\begin{equation}
\langle c^{\dagger}_{j\uparrow}(t)c_{i\uparrow}(0)\rangle\simeq 
\frac{e^{\pm i k_f x}}{[(x\pm v_st)(x\pm v_h t)]^{\frac 1 2}(x^2-v_h^2 
t^2)^{\frac 1{16}}},
\end{equation}
which can be easily shown to give a momentum distribution near
$k_f$ as
\begin{equation}
n(k_f)\sim n(k_f)-c |k-k_f|^{\frac 1 8} sgn (k-k_f).
\end{equation}
The lack of a finite jump at $k=k_f$ implies the vanishing spectral weight, i.e., 
$Z(E_f)=0$ at the Fermi points. 

One may also calculate the asymptotic spin-spin correlation function.
For example,
\begin{equation}\label{essc} 
\langle S_i ^+(t)S_j ^-(0)\rangle=\left\langle b_{i \uparrow}^+(t)
b_{j \uparrow}(0) b_{i \downarrow}(t)
b_{j \downarrow}^+(0)\right\rangle \left\langle e^{\pm i 2\Theta_i^h(t)}e^{\mp i2 \Theta_j^h(0)}\right\rangle .
\end{equation} 
The averages on the right-hand-side are also easily determined for the ground 
state of $H_{eff}$:\cite{weng1}
\begin{equation}
\left\langle b_{i \uparrow}^+(t)
b_{j \uparrow}(0)b_{i \downarrow}(t)
b_{j \downarrow}^+(0)\right\rangle \propto \frac {1}{(x^2-v_s^2t^2)^{1/2}},
\end{equation}
and 
\begin{equation}\label{essph}
\left\langle e^{\pm i 2\Theta_i^h(t)}e^{\mp i2 \Theta_j^h(0)}\right\rangle \propto \frac {\cos[\frac {\pi} {2a}(1-\delta)
x]}{(x^2-v_h^2t^2)^{1/4}}  .
\end{equation}
As similar asymptotic form can be found for  $\langle S_i ^z(t)S_j 
^z(0)\rangle$ if the ``squeezed spin chain'' effect is considered. Then,
\begin{equation}
\langle {\bf S}_i (t)\cdot {\bf S}_j (0)\rangle\propto   \frac {cos(2k_fx)}
{ (x^2-v_s^2t^2)^{1/2}(x^2-v_h^2t^2)^{1/4}} .     
\end{equation}

Therefore, the well-known asymptotic behavior of the single-electron Green's 
function and the spin-spin correlation function at $J\ll t$ is easily and  correctly reproduced here. The phase string contribution plays an essential
role in (\ref{egreen}) and (\ref{essc}). A lesson which we can learn from this is that even though the decomposition (\ref{e1d}) is 
mathematically equivalent to the conventional slave-fermion and slave-boson 
formalisms, it has the  advantage of explicitly tracking the phase string effect, 
so that such a nonlocal singular phase is not  lost when one makes 
some mean-field-type approximation to the Hamiltonian. Such a  phase string 
effect is really crucial,  due to its non-repairable nature, to  the long-time
and long-distance Luttinger-liquid behavior studied here. Thus if
one were to start from a {\it conventional} slave-particle scheme, a 
non-perturbative method beyond the mean-field theory must be employed in order to deal with the nonlocal phase string effect hidden in the Hamiltonian. In contrast, in the present phase-string formulation,
a  mean-field type treatment of the Hamiltonian gives reasonable results.
 
Finally, we make a remark about the physical meaning of the decomposition 
(\ref{e1d}). For a usual slave-particle decomposition,  
the quantum number (momentum) of the electron is a simple sum of the momenta of 
``spinon'' and ``holon'' constituents (due to the convolution relation). Here 
in the decomposition (\ref{e1d}) the phase string factor will ``shift'' the 
relation of the electron momentum to those of {\it true} spinon and 
holon  excitations in a nontrivial 
way,  which is actually equivalent to the information provided by the 
Bethe-ansatz solution.  In fact,  Ren and Anderson\cite{ra} have 
identified a similar effect (called a ``phase-shift'' by them)
based on the Bethe-ansatz solution.  

\subsection{2D example: nonlocal interactions}

In the last section, the phase string effect has been shown on the 1D example to
modify the long-distance and long-time electron correlations in a dramatic way. 
The 1D system is special in that the phase string only appears 
in correlation functions  through the decomposition (\ref{emutual}), but it does not show any direct effect in the Hamiltonian. Namely, the phase string 
effect can be ``gauged away'' in the Hamiltonian for an open-boundary 1D chain. 
This is consistent with the picture that spinon and holon as elementary excitations are decoupled, as indicated by either the exact solution\cite{lieb} or the analytic considerations in the  $J\rightarrow 0$ limit.\cite{weng1} However, for a 2D system such 
phase string effect can no longer be simply gauged away in the Hamiltonian. It is 
described by the lattice gauge fields $A_{ij}^f$ and $A_{ij}^h$ in the 
Hamiltonians  (\ref{et}) and (\ref{ej}), which are the Chern-Simons type  
gauge fields satisfying conditions (\ref{eflux1}) and (\ref{eflux2}), respectively. Therefore, even if there exists a spin-charge separation in
the 2D case, one still expects to  see nonlocal interactions between the spin and 
charge degrees of freedom, which may lead to anomalous transport and magnetic
phenomena.

In order to see the consequences of the gauge fields $A_{ij}^f$ and $A_{ij}^h$,
let us consider a mean-field type of approximation to $H_t$ and $H_J$ in 
(\ref{et}) and (\ref{ej}). This mean-field theory is similar to the one outlined above for the 1D case. But for the 2D case the 
mean-fields $\chi_0$ and $H_0$ should be defined with $A_{ij}^f$
and $A_{ij}^h$ incorporated as: $\chi_0=\frac 1 2 \sum_{\sigma}\left\langle e^{i\sigma 
A_{ij}^h} b_{i\sigma}^{\dagger}b_{j\sigma}\right\rangle$ and $H_0=\left\langle 
e^{i A_{ij}^f} h_{i}^{\dagger}h_{j}\right\rangle$. Correspondingly, the 
mean-field Hamiltonian $H_{eff}=H_h+H_s$ can be obtained similarly  to  the 1D 
case  (\ref{eh}) and (\ref{es}) as follows
\begin{equation}\label{ehh}
H_h=-t_h\sum_{\langle ij \rangle}\left(e^{iA_{ij}^f}\right)h^{\dagger}_i h_j + H.c.,
\end{equation}
and 
\begin{equation}\label{ess}
H_s=-J_s\sum_{\langle ij\rangle \sigma}\left(e^{i\sigma A_{ij}^h} \right) 
b_{i\sigma}^{\dagger}b_{j\sigma}+ H.c.
\end{equation}
This mean-field solution has been previously obtained based on the slave-boson
formalism,\cite{weng2} and the additional gauge fluctuations beyond the mean-field state
can be shown\cite{weng2} to be suppressed (gapped) at finite doping,  so that it is a real
spin-charge separation state, even though there exist  Chern-Simons 
(topological) fields $A_{ij}^f$ and $A_{ij}^h$ representing nonlocal 
scattering between the spin and charge degrees of freedom.

Let us first take a look at the charge degree of freedom. Due to the spin-charge 
separation (i.e.,  suppression of the gauge fluctuations), the charge response 
to external fields is entirely determined by the holon part described by $H_h$. 
According to (\ref{ehh}) as well as  (\ref{eflux1}), holons always see fictitious flux
tubes bound to spinons and quantized at $\pm \pi$, besides a lattice $\pi$-flux per plaquette,  as 
represented by $A_{ij}^f$.
It is important to note that the fluctuating part of  $A_{ij}^f$ not only
provides a scattering source in the long-wavelength, but also profoundly shapes
the coherent motion of holons.\cite{weng2} The latter effect is caused by strong short-range 
phase interference induced by the $\pm\pi$ flux quanta. This effect can be 
understood as the quantum interference of phase strings from different paths 
of holons. A semiclassical 
treatment of $H_h$ has been given in Ref. \onlinecite{weng2}, where the topological gauge
field $A_{ij}^h$ was shown to lead to  anomalous transport phenomena including 
linear-temperature resistivity,
a second scattering rate $\sim T^2$ in the Hall angle, a strong 
doping-dependence of thermoelectric power, etc., which gives a systematic
description of experimental measurements in the high-$T_c$ cuprates.     

The spin degree of freedom is also nontrivial here. As described by $H_s$ in 
(\ref{ess}), spinons see similar fictitious flux tubes bound to holons as 
represented by $A_{ij}^h$. It implies a strong frustration effect on the spin 
background induced by doping: each hole not only means a removal of a 
single spin, but also affects the rest of the spins nonlocally. This is a direct
consequence of the nonlocal phase string effect. Within this approximation, 
it has been found\cite{weng2} that the spin dynamics is  dramatically affected by the 
doping effect, including the emergence of low-lying doping-dependent magnetic
energy scales, non-Korringa behavior of the spin-lattice relaxation rate, etc.,
which qualitatively agrees with the anomalies found in the 
nuclear-magnetic-resonance (NMR) and neutron-scattering measurements of the 
high-$T_c$ cuprates. 

The above mean-field type theory has been previously developed by using the
slave-boson formalism.\cite{weng2} Here the phase string effect and its corresponding formalism
in Sec. III provide both physical and mathematical justifications for this 
approximate theory of the 2D $t-J$ model. Of course, a further improvement of
the theory based on the exact formulation given in Sec. III is desirable. For 
example,  the spin
degree of freedom as described by the mean-field Hamiltonian $H_s$ is still 
rather rough for short-range correlations. In particular, at half-filling where
the doping effect represented by $A_{ij}^h$ vanishes,  $H_s$ in (\ref{ess}) reduces to
a lattice model of a hard-core boson gas whose Bose-condensation gives rise to 
the long-range AF order. However, an accurate description of the 
antiferromagnetism in the ground state involves a RVB-type pairing of bosonic 
spins,\cite{liang} with the better mean-field solution at half-filling  known as the 
Schwinger-boson mean-field state\cite{schwinger} in which a pairing order parameter of spins is
used to describe short-range spin-spin correlations.  Based on the present 
formalism, it is not difficult to generalize the mean-field theory to 
incorporate such a short-range spin-spin correlation effect, which is important for
quantitatively explaining the experiments. A brief description of such a generalization has been reported in Ref.\onlinecite{weng4}, and a more detailed
version will be presented in follow-up papers.

\section{SUMMARY}

In the present paper, we have reexamined the problem of a hole moving
in an antiferromagnetic spin background and found rigorously that the
hole always acquires a nontrivial phase string at low energy. This 
phase string effect, particularly in 2D,  has been overlooked before, but its 
quantum interference effect can drastically change the hole's long-distance behavior. 
We have 
shown generally that  the spectral weight $Z$ must  vanish at the ground-state
energy due to such a phase string effect, which means that  the conventional 
perturbative description based on a quasipariticle picture should fail at
a sufficiently large distance in this system. The origin of this phase string
effect is related to the intrinsic competition between the superexchange and
hopping processes. Namely, the hopping of the hole displaces the spins  
in such a way that the spin-displacement (mismatch) cannot  
{\it completely} relax back via low-lying spin flips, and there is always a 
residual phase string left behind.

The phase string effect is not uniquely restricted to the one-hole
problem. It is also crucial at nonzero doping concentrations 
with or without a long-range order. The key issue is how 
one can mathematically describe the {\it long-distance} quantum effect of those
phase strings associated with the doped holes. Such an effect is hidden in the conventional 
slave-boson formalism (as a kind of sign problem).  And even though it shows up 
in a manifest way in the slave-fermion  formalism after the Marshall sign rule is 
included, its topological role as a Berry phase at large distances is still not explicitly
tracked in such a {\it local} model.  Thus, any local approximation
applied to those conventional  formalisms can easily damage the presumably crucial long-distant 
phase string effect and  may result in a   wrong physics in low-energy, long-wavelength regime. 

As discussed in the present paper,   the non-repairable phase string  on a closed-path is equivalent 
to a Berry phase.  It can be actually ``counted''  in terms of how many $\downarrow$ (or $\uparrow$) 
spins encountered by a given hole on its path. Thus such a Berry phase can be exactly tracked in the wavefunction.  Then,  based on a spin-hole basis with the built-in phase string effect, 
we obtain a new mathematical formulation for the $t-J$ model, in which the originally-hidden nonlocal phase string  effect is now explicitly represented in the Hamiltonian as {\it interacting effects} 
described by gauge fields with vorticities in the 2D case.  On the other hand, a {\it singular} part of the phase string effect is kept in the decomposition representation 
for the electron $c$-operator (i.e.,  in the wavefunctions) which is  crucial when one tries to 
calculate electron correlation functions (as shown in the 1D example).  

Another way to understand this new formulation is in terms of
so-called ``mutual statistics''. It has been pointed out that the phase string effect as a ``counting problem'' 
can be also  related to  the ``mutual statistics'' between  the charge and spin degrees of freedom, since each
step of hole hopping may be regarded as an exchange of a hole and a spin.  By using the composite representation of  the ``mutual statistics'' holon and spinon in the conventional bosonic description,  one 
can get the same formulation of the $t-J$ model in which the ``mutual statistics'' is  described by
long-range topological-type interactions (in 2D case) represented by nonlocal gauge fields. In contrast to
the fractional statistics among the same species, though, no explicit $T$- and $P$- violations are present in
this  ``mutual statistics'' or phase string description.

As an example, we have shown how the correct asymptotic behaviors of the 
single-electron Green's function and spin-spin correlation function can be
easily reproduced in the present scheme in the 1D  finite doping case. The present phase-string 
formulation proves to be very powerful in
dealing with this 1D problem, in contrast to difficulties associated with the  
conventional slave-particle formalisms. We have also discussed a 2D example by reproducing 
an approximate theory\cite{weng2} which gives a systematic description of the 
anomalous transport and magnetic properties in the high-$T_c$ cuprates. Such a theory was
previously developed\cite{weng2} based on the slave-boson scheme with an optimization
procedure (known as flux-binding) at small doping, with the key mechanism being  
topological gauge-field interactions between spinons and holons. The present
phase-string theory lays a solid  foundation for such a mechanism, and provides a basis for the further improvement of the generalized mean-field theory and for a more quantitative comparison with the experiments. We  will address these issues in follow-up papers.

\acknowledgments
The authors acknowledge helpful discussions with T. K. Lee, B. Friedman, Y. S. Wu, O. Starykh, 
D.K.K. Lee, N. Nagaosa, P.A. Lee,  G. Baskaran,  N.P. Ong, P.W. Anderson, and B. Marston. 
We would like to especially thank D. Frenkel and T.M. Hong for their critical readings of the manuscript.  
The present work is supported by  Texas Advanced Research Program  under Grant No. 3652182, a 
grant from the Robert A. Welch foundation, and 
by Texas Center for Superconductivity at University of Houston.\\ 
 
* Permanent address: Department of Physics, University of Science and 
Technology of China, Hefei, Anhui 230026, China.


\begin{references}
\bibitem{remark1} In a semiclassical description, the $U(1)$ phase effect is
represented by transverse spin components which exhibit a large-scale 
dipolar-like structure as discussed in Ref. \onlinecite{shraiman}.
\bibitem{remark} Here one should be cautious  
about some special cases where $\rho(E_G)=0$.  For example, it occurs at a  
dimensionality higher than 2D for a parabolic quasiparticle spectrum, and in 2D 
for a  massless Dirac spectrum $E_k$.
\bibitem{remark3} A factor $e^{\pm i \frac {1-\sigma}{2} \hat{N}_h}=(\sigma)^{\hat{N}_h}$ ($\hat{N}_h=\sum_l n^h_l$)
is omitted in (\ref{emutual}) for simplicity, which  has 
no physical effect except for maintaining anticommutation relation 
$\{c_{i\uparrow},c_{j\downarrow}\}= 0$ ($i\neq j$), etc., for opposite spins
(Klein factor).
\end{references}
\end{document}